\documentclass[sigconf,nonacm]{acmart}

\AtBeginDocument{%
  }

\usepackage{comment}

\begin{document}

\title[Data Center Cooling Choices Shift Water Impacts Across the Grid]{Data Center Cooling Choices Shift Water Impacts Across the Grid: An Integrated Water-Energy Model for Sustainable Data Center Development}

\author{Garrett Alston}
\email{galston@umich.edu}
\affiliation{%
  \institution{Department of Civil and Environmental Engineering, University of Michigan}
  \city{Ann Arbor}
  \state{Michigan}
  \country{USA}
}

\author{Nancy Love}
\email{nglove@umich.edu}
\affiliation{%
  \institution{Department of Civil and Environmental Engineering, University of Michigan}
  \city{Ann Arbor}
  \state{Michigan}
  \country{USA}
}

\author{Rabab Haider}
\email{rababh@umich.edu}
\affiliation{%
  \institution{Department of Civil and Environmental Engineering, University of Michigan}
  \city{Ann Arbor}
  \state{Michigan}
  \country{USA}
}

\renewcommand{\shortauthors}{Alston et al.}

\begin{abstract}
Data centers are being developed at an unprecedented pace, yet their energy and water impacts, and the spatial and temporal distribution of these impacts, remain poorly characterized. Data centers consume water for cooling (direct) and through electricity generation (indirect). Decisions on siting and cooling technology result in water--energy trade-offs that extend impacts beyond the facility’s location. Existing assessment frameworks rely on facility efficiency metrics and average grid water intensity factors, suppressing the temporal impacts of data center load and generation availability. They also attribute indirect consumption to the facility's location rather than to the generators (and corresponding hydrologic regions) that respond to the added load, misattributing the spatial impacts of data center loads. To close this gap, we develop a computational model of the data center--energy--water nexus that links facility cooling and electricity demand with hourly economic dispatch, generator-level water consumption, and monthly subbasin depletion. Built on open-source data, the model resolves where and when water is consumed, and where this consumption compounds existing water risk or creates new risk. We apply the model to cooling technology selection and to proposed developments in the state of Michigan. Air-cooled data centers nearly halve total water consumption relative to evaporative cooling, but increase electricity demand and raise indirect water consumption by roughly one-third, shifting the water footprint from the facility to generators on the grid. Indirect water intensity also varies by approximately 50\% across months with the generators supplying the added load. Mapping these changes to subbasins reveals depletion increases beyond the data center sites, in regions that facility-level reporting may overlook. These results show that data center water and energy impacts cannot be assessed in isolation, motivating the need for integrated modeling to inform siting, design, and reporting practices.
\end{abstract}


\maketitle

\section{Introduction}
The exponential growth of data centers is increasing electricity demand worldwide, with continued expansion and increased consumption expected as computing workloads and digital services grow \cite{iea25}. The United States accounts for the largest share of global data center electricity consumption \cite{iea25}, with U.S. facilities consuming an estimated 192 TWh in 2024, accounting for 4.7\% of national electricity consumption \cite{smith26}. Data centers also consume substantial quantities of water: \emph{Scope 1 or direct} consumption is the on-site water used at the facility, primarily for cooling; and \emph{Scope 2 or indirect} consumption is the off-site water used at the electricity generation facilities to power the data center \cite{mytton21}. Critically, Scope 2 consumption links energy and water impacts: decisions around facility siting, cooling technology selection, and energy supply result in tradeoffs across the water--energy nexus, and across spatial and temporal dimensions.

Cooling configurations that reduce direct water consumption may increase electricity demand and the indirect water consumption associated with electricity generation. Evaporative cooling systems consume relatively large quantities of direct water and require less electricity to operate, whereas non-evaporative systems consume relatively small quantities of direct water but require more electricity \cite{karimi25}. The resulting environmental impact of a data center cannot be inferred from metrics reporting direct water consumption alone. The proposed integrated data center--energy--water model combines three metrics. Power usage effectiveness (PUE) relates total facility electricity to information technology (IT) electricity used by servers, storage, and networking equipment. Water usage effectiveness (WUE) relates site water use to IT electricity. Water consumption intensity (WCI) relates the volume of water consumed by a generator per unit of electricity generated. Together, these three metrics determine the water impact of a data center facility: the cooling configuration determines the WUE and PUE of the facility, and can shift water consumption between the data center and the generators responding to its electricity demand, which are represented by the WCI.

Importantly, these impacts also vary temporally and spatially. Cooling performance changes with computing load and climate conditions, while electricity demand, renewable availability, and generator dispatch vary throughout the year. Direct water consumption occurs at the data center, while indirect water consumption occurs at the generators responding to its demand, so the water impacts can occur across different hydrologic regions \cite{siddik24}. Assessing water risk requires a consistent way to measure water use and assess subsequent impacts. \emph{Withdrawal} is the total volume removed from a water source, whereas \emph{consumption} is the portion of that withdrawal not returned to the immediate water environment \cite{macknick11}. These two quantities support different risk indicators. \emph{Water stress} is commonly represented by the ratio of water withdrawals to available water, whereas \emph{water depletion} is the ratio of consumptive water use to renewable available water \cite{brauman16}. Because depletion accounts for return flows and captures the share of available water that is actually consumed, this study uses a depletion-based indicator to evaluate water risk. Therefore, the same increase in consumption can result in different water risk values, depending on existing consumption and available water where it occurs.

\subsection{Related Works and Research Gaps}

Data center water--energy assessments have increasingly expanded the system boundary to include both facility and electricity generation water use. Early approaches estimated indirect water consumption by applying grid-averaged water consumption factors to facility electricity use based on regional power mixes \cite{greengrid11,ristic15}. Later studies incorporated differences in cooling technology, climate, and regional water availability \cite{siddik21,lei23,karimi25} and, more recently, spatial and temporal variation in electricity generation and hydrologic conditions \cite{wu25,siddik24}. Across all of these studies, however, the indirect water consumption calculation retains the same fundamental limitation: it uses the water consumption intensity of an averaged generation mix rather than identifying the generators whose output actually changes in response to that demand.

This limitation persists even at higher temporal resolutions. Existing assessments assign added demand the water consumption intensity of generation already occurring in the region and period analyzed \cite{siddik21,siddik24}, assuming generators increase their output in proportion to the observed mix. This assumption simplifies the true grid response. These approaches can estimate the water consumption when the generation mix is fairly uniform (e.g., all natural gas, or all hydro), a limiting assumption for most real-world systems \cite{Miller_2022_hourlycarbonaccounting}. 

This limitation is even more pronounced for spatial impacts. Aggregating generator water consumption into regional electricity intensities does not preserve the locations of responding generators, so it cannot assign incremental consumption to the hydrologic regions where it actually occurs. While direct cooling consumption falls within the facility's hydrologic region, the generators responding to its electricity demand may be located and consume water in several other regions \cite{ristic15,siddik21}. Therefore, an assessment based on the facility location \cite{wu25} or an aggregated electricity region \cite{siddik24} may estimate the total indirect consumption volume while misidentifying the regions in which the consumption actually occurs.

\subsection{Contributions}
Thus, existing approaches cannot resolve which generators respond to added demand, nor the hydrologic regions in which that response's consumption and depletion occur. To bridge this gap, we present an integrated water--energy model that connects facility operations, grid dispatch, and water subbasin conditions to evaluate the water and energy impacts of data center development. We then use the integrated model to evaluate the impacts of facility siting, cooling configuration, and computing workload on water risk, which we define as a depletion-based metric at the subbasin level. Our model is built using publicly available real-world data, and used to simulate data center growth in the state of Michigan. Our case studies show:
\begin{itemize}
    \item Reducing direct cooling consumption can increase electricity demand and indirect consumption, highlighting why cooling choices must be evaluated across both the facility and the specific generators responding to its load.
    \item Indirect water intensity varies across months as the responding generators change, revealing temporal differences that are obscured by applying a fixed regional electricity water-intensity factor.
    \item Assigning direct and indirect consumption to their respective subbasins reveals depletion increases beyond the data center sites that would be missed by evaluating both components using only the facility's hydrologic boundary.
\end{itemize}

\section{An Integrated Water--Energy Model}
The integrated model is composed of three systems:
\begin{itemize}
    \item \textbf{DCF System: Data center facility design and operations.} Models data center siting (lat/long coordinates), IT capacity or total facility capacity, hourly computing load, cooling configuration, power usage effectiveness (PUE), and water usage effectiveness (WUE). This system calculates the Scope 1 water consumption from cooling using the facility WUE.
    \item \textbf{Energy System: Regional power grid operations and generator dispatch.} Models the power grid infrastructure (substations and transmission lines), hourly regional electricity demand, and generators. Individual generators' operating costs and capacity are used in the economic dispatch optimization. This system calculates Scope 2 water consumption from electricity generation using generator water consumption intensities. The Energy System is coupled with the DCF system through the facility PUE.
    \item \textbf{Water System: Consumption and depletion for hydrologic regions.} Models the hydrologic region boundaries, and monthly baseline water availability and consumption. This system calculates the water risk using a depletion-based metric at the subbasin level. The Water System is coupled with the other two through the water depletion from cooling (Scope 1, DCF System) and depletion from generation (Scope 2, Energy System).   
\end{itemize}
Throughout this paper, indirect consumption refers only to electricity generation. These pathways do not constitute a complete life-cycle analysis of the water footprint, as water embodied in facility construction, server and equipment manufacturing, and other supply chain activities is considered Scope 3 use and is outside the model boundary of this study \cite{alissa25}.

\begin{figure}[t]
    \centering
    \includegraphics[
        width=\columnwidth,
        trim={2.5cm 2.0cm 2.5cm 1.8cm},
        clip
    ]{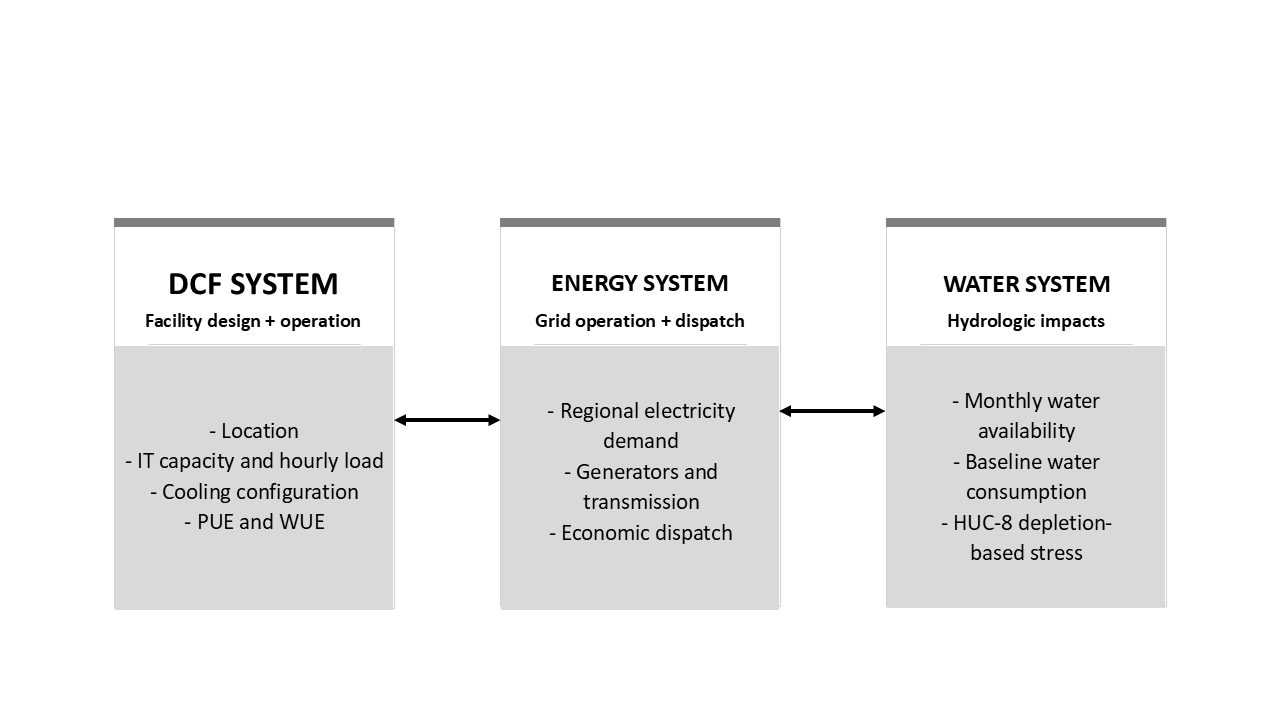}
  \caption{Integrated water--energy model is composed of three systems: data center facility, energy, and water systems}
  \Description{Three connected systems represent data center design, grid operations and dispatch, and hydrologic conditions and depletion, with arrows showing system coupling.}
\end{figure}

\subsection{DCF System model}

Each facility $f\in\mathcal{F}$ is specified by its coordinates, capacity, electricity load profile, and cooling configuration. The facility capacity can be defined as either IT demand or total facility load. IT load includes servers, storage, and networking equipment, while total facility load includes IT demand as well as cooling, power conversion, lighting, and other supporting systems \cite{sun21}. We use PUE and WUE metrics to determine facility electricity and water use for different cooling configurations. These metrics primarily remain fixed; however, when modeling cooling configurations with multiple operating modes, they vary over time with the selected mode.

\subsubsection{IT-load representation}
Facility electricity for time interval $t$ is calculated from the IT load $P_{\mathrm{IT},f}$, using the facility's reported PUE:
\begin{equation}
E_{\mathrm{facility},f}
=
P_{\mathrm{IT},f}\times\Delta t\,\times\mathrm{PUE}_{f},
\label{eq:facility_energy}
\end{equation}
where $P_{\mathrm{IT},f}$ is in MW, $\Delta t$ is the interval duration in hours, and $E_{\mathrm{facility},f}$ is in MWh.

\subsubsection{Facility-load representation}
In this configuration, the total facility load is
\begin{equation}
E_{\mathrm{facility},f} = P_{\mathrm{facility},f}\times\Delta t
\label{eq:profiled_facility_energy}
\end{equation}
and the IT load is calculated from the PUE as
\begin{equation}
E_{\mathrm{IT},f} = \frac{E_{\mathrm{facility},f}}{\mathrm{PUE}_{f}}.
\label{eq:it_energy}
\end{equation}

\subsubsection{Scope 1 Water Consumption}
Under both load representations, the model treats WUE as a consumptive cooling water intensity, denoted as $\mathrm{WUE}_{f}$ in L/kWh of IT electricity. Direct consumption is
\begin{equation}
W_{\mathrm{direct},f}
=
E_{\mathrm{IT},f}\times\,\mathrm{WUE}_{f},
\label{eq:direct_water}
\end{equation}
where $E_{\mathrm{IT},f}$ is in MWh and $W_{\mathrm{direct},f}$ is in m$^3$.

\subsubsection{Industry data: PUE and WUE metrics}
The PUE and WUE metrics are taken from the 2024 Department of Energy report, the \textit{United States Data Center Energy Usage Report} \cite{shehabi24}. The report provides simulated ranges of annualized PUE and WUE across data center space types, IT cooling systems, and heat rejection systems. These ranges come from thermodynamic-based models evaluated across surveyed data center operating and climate conditions \cite{shehabi24}.

It is important to note the distinction between the IT cooling system (removes heat from the servers) and the heat rejection system (removes heat from the facility). For example, describing a system as ``liquid cooled'' identifies how heat is removed from the servers but does not establish whether heat is then rejected from the facility through a cooling tower, fan towers, or other heat rejection methods. The overall facility cooling configuration in our model refers to the combination of technologies selected for IT cooling and facility heat rejection.

\begin{figure}[h]
  \centering
  \includegraphics[width=\linewidth]{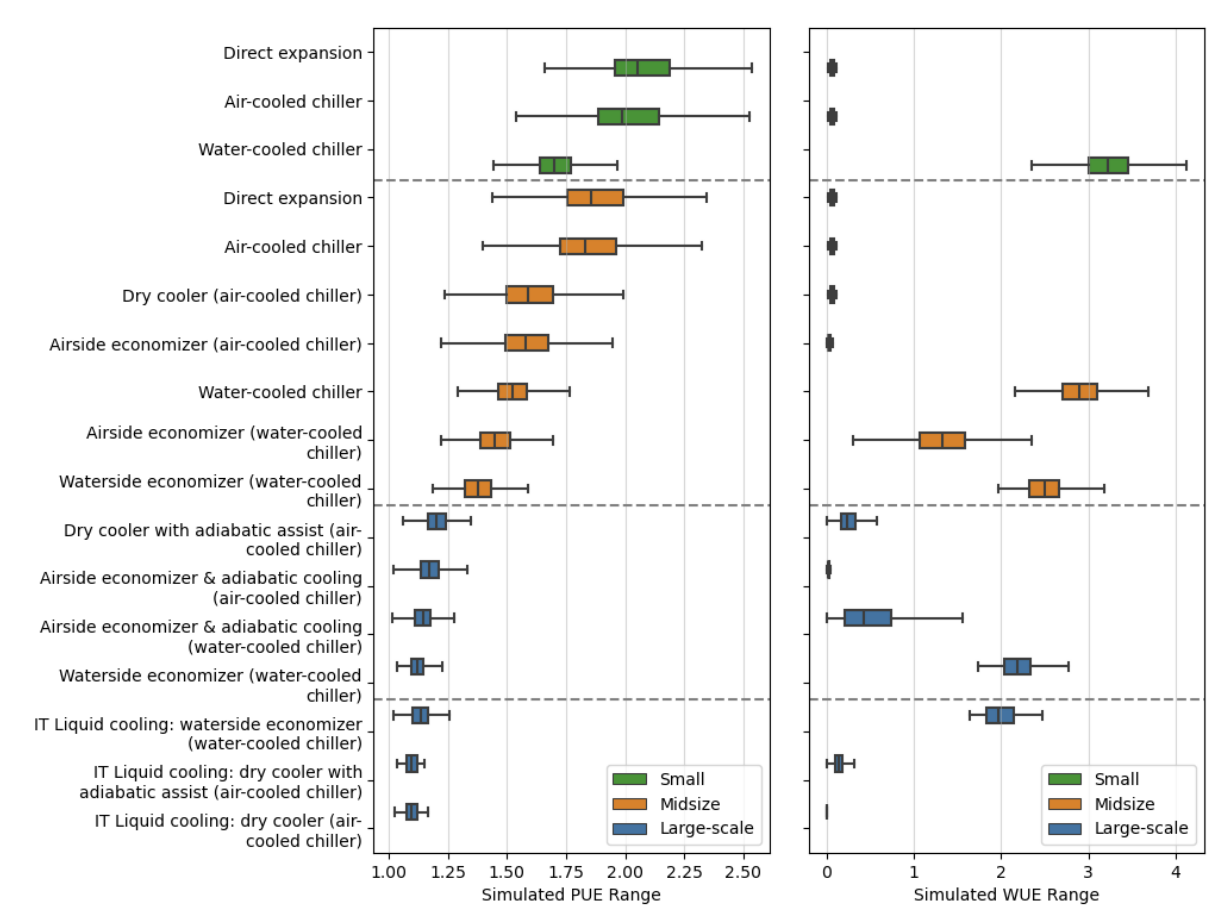}
  \caption{Data from the 2024 DOE report on data center energy usage, showing PUE and WUE ranges by cooling configuration and data center size \cite{shehabi24}(see Figure 4.4)}
  \Description{Two panels show simulated PUE ranges on the left and WUE ranges on the right for different cooling configurations. Green, orange, and blue indicate small, midsize, and large-scale data centers, respectively.}
  \label{fig:doe_pue_wue_ranges}
\end{figure}

\subsection{Energy System model}
The transmission grid is modeled as a graph consisting of electrical buses (graph nodes) $\mathcal{N}=\{1,...,n\}$ and transmission lines (graph edges) $\mathcal{L}=\{1,...,m\}$. Transmission line limits are given by $\overline{F}\in\mathbb{R}^m$. Generators are represented by a fleet of $k$ generators with capacity limits $p\in[\underline{P},\overline{P}]$, where $\underline{P}, \overline{P}\in\mathbb{R}^k_{\geq0}$ and linear cost curves of the form $cp$, where $c\in\mathbb{R}^k_{+}$ is the marginal cost of generation and $p$ is the dispatch setpoint. Generators are mapped to buses using matrix $\mathcal{G}\in\mathbb{R}^{n\times k}$, where $\mathcal{G}_{ij}=1$ if generator $j$ injects power at bus $i$. Electric load is given by the demand vector $d\in\mathbb{R}^n_{\geq0}$. 

\subsubsection{Generator data}
\label{gen_data}
The generators and their nameplate capacities are taken from the U.S. Energy Information Administration (EIA) Form 860, which reports power plant and generator data in the U.S. \cite{eia24a}. Individual generators at the same plant are grouped by energy technology and represented as a single generator $g$ in the grid model with their summed nameplate capacities. Each facility and generator is assigned to an electrical bus: facility (model input) and generator coordinates (form EIA-860) are mapped to the closest bus using the Haversine distance between the two points. The sets $\mathcal{F}_i$ and $\mathcal{G}_i$ denote the facilities and generators assigned to bus $i$, respectively.

\subsubsection{Generator availability}
Dispatchable generators (e.g., thermal generation and pumped hydro) are assumed to be dispatched up to their nameplate capacity. Intermittent renewable generation (e.g., wind and solar) is modeled with hourly availability data, as a fraction of their nameplate capacity. Wind is modeled using county-level hourly capacity factors from the Regional Energy Deployment System (ReEDS) renewable dataset \cite{sergi25}. Solar is modeled by matching solar plants to the nearest National Solar Radiation Database (NSRDB) 4 km resolution location, with hourly availability approximated from normalized global horizontal irradiance \cite{sengupta18}. All generating technologies are assumed to have no minimum generation requirements, with the exception of nuclear, which is assigned a 95\% minimum load requirement to represent sustained operations. Wind and solar generation can be curtailed if available generation exceeds system load or transmission congestion limits the deliverability to load sites. 

\subsubsection{Grid operations and generator dispatch}
Generator dispatch is modeled using the structure of U.S. electricity markets, using a linearized optimal power flow model. Total system generating cost is minimized, subject to network constraints (Kirchhoff’s Voltage and Current Laws, thermal line limits). The optimization output is the set of generators that are dispatched to supply the system load, their power setpoints, and total system costs. We use the open-source PyPSA Python package to model the grid, generators, and economic dispatch problem, which is solved with the HiGHS optimization solver \cite{brown18}.

\subsubsection{Generator water consumption}
\label{gen_water_consum}

The EIA provides monthly water consumption intensities for generators in the thermoelectric cooling water dataset, derived from EIA-860 forms \cite{eia24a,eia24b}. Consumption per unit of generation is expressed in U.S. gal/MWh and denoted as $\mathrm{WCI}_{g,m}$, and reported for generators with capacities above 100 MW. To address partial or missing data for generators, the monthly $\mathrm{WCI}_{g,m}$ values are assigned as follows:
\begin{enumerate}
\item Where monthly data is reported for a generator, the generator-specific values are taken directly from the EIA dataset.
\item If no information is available for a specific generator in a given month, but data is available for other generators of the same technology during that month, it is assigned the monthly median for its corresponding generation technology. The monthly median is calculated per technology, using all available generator data from the EIA dataset.
\item If no information is available for a specific generator in a given month and no data is available for other generators of the same technology during that month, it is assigned the annual average of available monthly technology medians for its corresponding generation technology.
\item If no information is available for any generator of a specific technology type, the WCI values are manually assigned from other sources.
\end{enumerate}
 
\noindent The water consumption for a generator $g$ in month $m$ is
\begin{equation}
W_{\mathrm{generation},g,m}
=
\gamma\,\times \mathrm{WCI}_{g,m}\times
\sum_{t\in m}p_{g,t}\Delta t,
\label{eq:generator_water}
\end{equation}
where $\gamma$ converts water volumes from U.S. gallons to m$^3$, $\Delta t$ is the interval duration in hours, and $p_{g,t}$ is the power dispatch of the generator $g$ at time $t$ in MW. This calculation links the Energy System to the Water System.

\subsection{Water System model}

The Water System uses Hydrologic Unit Codes (HUCs) to define the boundaries of drainage areas. HUC-8 units represent subbasins, while HUC-12 units represent smaller subwatersheds. Subbasins are used as the reporting scale because they balance spatial granularity with regional interpretability. Their scale is coarse enough to avoid reporting results across hundreds of individual HUC-12 units, while remaining sufficiently granular to preserve spatial differences in the HUC-12 data. Note that the proposed model can be resolved at other hydrologic scales, if sufficient data is available at that spatial scale. Each facility and generator is assigned to a subbasin: facility (model input) and generator coordinates (form EIA-860) are mapped to the containing subbasin polygons. The sets $\mathcal{F}_h$ and $\mathcal{G}_h$ denote the facilities and generators assigned to subbasin $h$, respectively. It should be noted that this mapping is separate from assignment to electrical buses -- because facilities sharing a bus (subbasin) do not necessarily share a subbasin (bus).

Monthly baseline conditions come from the U.S. Geological Survey (USGS) National Water Availability Assessment outputs \cite{usgs25}. The HUC-12 USGS consumption and availability data, where water quantities are reported as depths in mm/month, are aggregated to HUC-8 using area-weighted means. Consumption includes water consumed by public supply, thermoelectric generation, and crop irrigation, and is cumulative over the local and upstream drainage area. Availability represents water exiting a HUC-12 after consumptive losses, and represents the remaining water available for use by the facility and generators supplying the additional load. Let $A_{h,m}$ and $C_{h,m}$ denote the availability and consumption depths for month $m$ at the HUC-8 unit. The gross supply for a subbasin $h$ is
\begin{equation}
S_{h,m}=A_{h,m}+C_{h,m}.
\label{eq:gross_supply}
\end{equation}
The resulting supply and consumption depths provide the inputs for the final depletion-based water risk calculation.

\subsection{Water Risk}
The integrated water--energy model evaluates the impact of data center $f$ by comparing the generation dispatch and water consumption to a baseline without any facility additions. The monthly change in generation from the facility addition compared to the baseline is
\begin{equation}
\Delta G_{g,m}
=
\sum_{t\in m}
\left(
p^{\mathrm{DCF}}_{g,t}
-
p^{\mathrm{baseline}}_{g,t}
\right)\Delta t,
\label{eq:dispatch_change}
\end{equation}
where $\Delta G_{g,m}$ is measured in MWh, and 
$p^{\mathrm{DCF}}_{g,t}$ and $p^{\mathrm{baseline}}_{g,t}$ are the power output of generator $g$ during time interval $t$ with and without the data center, respectively. The corresponding change in generator water consumption is
\begin{equation}
W_{\mathrm{indirect},g,m}
=
\gamma\ \times\Delta G_{g,m}\times\mathrm{WCI}_{g,m}.
\label{eq:indirect_water}
\end{equation}
Note that only the change in generator water consumption is added to the hydrologic baseline because the USGS baseline $C_{h,m}$ already includes thermoelectric water use. 

Direct and indirect water consumption are aggregated to subbasin $h$, and converted from volume (in $m^3$) to depth quantities (in mm) using the subbasin area $A_h$ as
\begin{align}
W_{\mathrm{direct},h,m} &= \frac{1000}{A_h} \sum_{f\in\mathcal{F}_h}\sum_{t\in m} W_{\mathrm{direct},f,t}
\label{eq:huc_water_aggregation_direct} \\
W_{\mathrm{indirect},h,m} &= \frac{1000}{A_h}\sum_{g\in\mathcal{G}_h}W_{\mathrm{indirect},g,m}.
\label{eq:huc_water_aggregation_indirect}
\end{align}

\noindent The water depletion for the baseline and facility addition are
\begin{align}
D^{\mathrm{baseline}}_{h,m} &= \frac{C_{h,m}}{S_{h,m}},
\label{eq:baseline_stress} \\
D^{\mathrm{DCF}}_{h,m} &= \frac{C_{h,m} + W_{\mathrm{direct},h,m} + W_{\mathrm{indirect},h,m}
}{S_{h,m}}.
\label{eq:dc_stress}
\end{align}
and the absolute change in depletion with the facility addition, in percentage points (pp), is:
\begin{equation}
\Delta D_{h,m} = \left(D^{\mathrm{DCF}}_{h,m}-D^{\mathrm{baseline}}_{h,m}\right)\times100.
\label{eq:depletion_change}
\end{equation}

Throughout the calculations, the direct and indirect terms remain assigned to their respective subbasins, capturing the full spatial impacts of facility addition beyond the specific site location. 

\section{Case Study}

To demonstrate the framework, the model is implemented using Michigan-specific data and applied to two scenarios that illustrate its ability to evaluate different elements of data center development and its impacts.

\subsection{Michigan Model Implementation}

\subsubsection{Grid data:}
Michigan's electricity system is represented using four regions, each represented as an electrical bus in $\mathcal{N}$: Southeast, Central, West, and North Michigan. Four transmission corridors connect Southeast to Central, Southeast to West, Central to West, and Central to North Michigan. Each corridor has a thermal limit $\overline{F}=6,000$ MW, selected heuristically to remain well above expected interregional power transfers and thereby prevent transmission congestion from dominating the dispatch results. Individual transmission lines are assigned line reactances of 0.10 $\Omega$ and line resistances of 0.01 $\Omega$ as heuristic parameters, consistent with values used in PyPSA’s network examples \cite{PyPSA}.

\subsubsection{Generator data:}
Michigan has 32.4 GW of installed generation capacity based on the EIA-860 data \cite{eia24a}. Natural gas accounts for 44.6\% of capacity, followed by coal (19.5\%), wind (11.7\%), nuclear (10.8\%), hydroelectric generation (7.2\%), solar (3.6\%), biomass (1.5\%), petroleum (1.3\%), and battery storage (less than 0.1\%). Based on the marginal costs assigned to each generating technology, detailed in Appendix \ref{app:mi_gen_data}, the economic dispatch model generally prioritizes dispatching wind and solar, followed by hydroelectric generation, nuclear and natural gas, coal, biomass, and petroleum.

The monthly water consumption intensities are assigned based on the procedure described in Section \ref{gen_water_consum} (see Appendix~\ref{app:mi_gen_data} for annualized WCI). The remaining technologies without data reported in the EIA dataset are assigned the fixed WCI values shown in Table\ref{tab:manual_wci}. Specific technologies present in Michigan's generation mix are treated as follows. Petroleum coke plants operate similarly to steam coal plants \cite{mhi07petcoke,eia20fuelconversion}, and are assigned the median WCI of steam coal reported in a study by the National Laboratory of the Rockies \cite{macknick11}. Hydroelectric generation (conventional and pumped storage) is assumed to have a WCI of zero. Many studies use high water consumption values, attributed to evapotranspiration from their reservoirs \cite{macknick11}. However, WCI is the rate of water consumption per unit of power produced; the operation of a hydroelectric plant does not significantly change the evapotranspiration of the reservoir, and water is not otherwise consumed in hydroelectric power generation. The WCI metric is more accurate in capturing the change in resource operations due to an addition of load.

\subsubsection{Load data:}
Electric load is constructed from the 2023 hourly county electricity profiles from the Open Energy Data Initiative's demand dataset \cite{obika25}. Each county is assigned to the nearest bus based on the Haversine distance from the county's centroid. The aggregated hourly electricity demand of all counties mapped to bus $i$ represents the demand $d_i$.

\begin{table}[t]
  \caption{WCI values for generating technologies not found in EIA dataset.}
  \label{tab:manual_wci}
  \centering
  \begin{minipage}{\linewidth}
  \small
  \begin{tabular}{@{}p{0.68\columnwidth}r@{}}
    \toprule
    Generating technology and source & WCI (gal/MWh) \\
    \midrule
    Batteries \footnotemark[1] & 0 \\
    Conventional hydroelectric \footnotemark[1] & 0 \\
    Hydroelectric pumped storage \footnotemark[1] & 0 \\
    Landfill gas \cite{macknick11} & 35 \\
    Natural-gas combustion turbine \cite{doe02} & 0 \\
    Natural-gas internal-combustion engine \cite{epa17chp} & 0 \\
    Onshore wind \cite{macknick11} & 0 \\
    Petroleum coke \cite{macknick11,mhi07petcoke,eia20fuelconversion} & 687 \\
    Solar photovoltaic \cite{macknick11} & 26 \\
    \bottomrule
  \end{tabular}
  \textsuperscript{1}{Values assumed based on technology operations}
  \end{minipage}
\end{table}

\subsection{Scenario 1: Evaluating Water--Energy Tradeoffs from Cooling Configurations}
Scenario 1 evaluates three cooling configurations at a hypothetical data center facility in Oakland County, Southeast Michigan, which presently has the highest concentration of operating data centers in the state \cite{banthia26}. The facility has a 200 MW IT load, defined using the IT-load representation of \eqref{eq:facility_energy}. Under this formulation, the cooling configuration impacts both direct water consumption (via WUE) and the facility electricity demand (via PUE), to directly evaluate water--energy tradeoffs and its spatial impacts. 

Three cooling configurations are evaluated: a water-cooled chiller, an air-cooled chiller, and a hybrid airside economizer paired with a water-cooled chiller. The water- and air-cooled chiller configurations each rely on a single cooling technology and maintain operations throughout the year. For both configurations, the PUE and WUE inputs are selected from the upper endpoints of their respective midsize data center performance ranges \cite{shehabi24}, representing worst-case configurations. 
The hybrid configuration switches between air-cooled and water-cooled configurations throughout the year. For the PUE and WUE metrics, the lower endpoints represent the air-cooled mode because it uses less energy (free/passive cooling) and less water (non-evaporative), and the upper endpoints represent the water-cooled mode because it uses more energy (from pumping and cooling towers) and more water (evaporative).

The daily operating mode is determined based on the daily average temperature and relative humidity of the site. Data from the 2023 National Solar Radiation Database (NSRDB) are taken from the grid point nearest the data center site using Haversine distance \cite{sengupta18}. The air-cooled mode (economizer) is selected when the mean temperature is at or below 27$^{\circ}$C and mean relative humidity is at or below 70\%. If either threshold is exceeded, the water-cooled mode (chiller) is selected. These thresholds correspond to the maximum recommended dry-bulb temperature and relative humidity of air for cooling IT technology in the Federal Energy Management Program’s \textit{Best Practices Guide for Energy-Efficient Data Center Design} \cite{vangeet24}.

\subsection{Scenario 2: Proposed Michigan Data Center Build Out}

Scenario 2 evaluates three proposed hyperscale data center developments in Michigan using information collected through FracTracker’s data center tracking tool \cite{fractracker26}. The tracker reports maximum facility demands of 2,700 MW, 1,400 MW, and 525 MW for the three selected facilities, resulting in a combined maximum demand of 4,625 MW. Although the capacity data from real proposed developments are used, the facility configurations are not known. All three developments are hyperscale data centers, likely for AI training workloads. These configurations typically use IT liquid cooling at the server side, but may use various heat rejection systems. This scenario assumes all three have IT-liquid cooling with a waterside economizer and water-cooled chiller, using the respective median values for PUE and WUE \cite{shehabi24}.

The facility-load representation of \eqref{eq:it_energy} is used, where each facility's maximum demand is scaled using the hourly synthetic load profile \cite{bertolacini26} and PUE is only used to calculate IT electricity for direct water consumption:
\begin{equation}
E_{\mathrm{facility},f}
=
P_{\mathrm{facility},f}\times L_{f,t}\times\Delta t,
\label{eq:scenario2_profiled_facility_energy}
\end{equation}
where $L_{f,t}$ is the fraction of maximum facility demand during interval $t$. PUE is then used only to calculate IT electricity using \eqref{eq:it_energy}, which determines direct water consumption through \eqref{eq:direct_water}.

To evaluate the impacts of not only the magnitude of added load, but also its location, two siting cases are simulated. The base case places all three facilities at their currently proposed locations in Southeast Michigan, representing concentrated development. The alternate case places the same facility loads at hypothetical locations in West, Southeast, and Central Michigan. These two locational cases could represent different data center compute: inference tasks that are latency sensitive may locate closer to users in urban environments, while AI training tasks that are less latency sensitive may locate further from users. Since the three facilities have the same cooling configuration and follow the same load profile, the total direct water consumption and electric load will be the same across both cases. However, the location of direct water consumption will vary based on the location of the facility. The indirect water consumption quantities and locations may also vary, as the load interacts with both generation and transmission capacity. 

Note that this scenario is an illustrative analysis of how significant data center additions can impact water and energy systems, and does not serve as an assessment of the impacts of any specific project.

\section{Results and Discussion}
Prior water--energy assessments of data centers have relied on facility efficiency metrics and average grid water intensity factors to estimate off-site consumption \cite{ristic15,lei23,lei25,shehabi24}. Because these approaches do not identify which generators respond to an incremental load, when that response occurs, or where the resulting consumption falls relative to existing water conditions, they suppress the temporal impacts and misattribute the spatial impacts of data center loads. Applying our integrated water--energy model to the Michigan case studies reveals five critical findings that these simplified approaches cannot capture:
\begin{enumerate}
    \item Cooling configuration results in a water--energy tradeoff
    \item Average grid water intensities cannot capture the true indirect water impacts
    \item Data center impacts extend beyond the facility site
    \item Annualized reporting obscures the underlying system conditions
    \item Water consumption magnitude alone cannot determine water risk
\end{enumerate}
These findings are discussed next, drawing from the results of both scenarios. Our findings have direct consequences for how the water and energy footprint of data center developments should be assessed, sited, and reported.

\subsection{Cooling Configuration Results in a Water--Energy Tradeoff} \label{sec:result:tradeoff}

Scenario 1 compares three cooling configurations, tracking the direct and indirect water consumption over a year of operations. Generally, configurations with lower direct water consumption (e.g., air-cooled) reduce total modeled water consumption, while increasing the relative contribution of indirect water use.

Table~\ref{tab:Ann_E_and_W} shows this tradeoff most clearly between the water-cooled and air-cooled configurations. Moving from the water-cooled chiller to the air-cooled chiller reduced annual direct water consumption by 97\%, from 6.45 to 0.21 million m$^3$, while increasing facility electricity demand by 32\%, from 3.08 to 4.06 TWh. This additional electricity demand increased indirect water consumption by 32\%, from 4.48 to 5.91 million m$^3$. Despite this increase, total water consumption decreased when moving to the air-cooled chiller by 44\%, from 10.93 to 6.12 million m$^3$. As a result, indirect water increased from 41\% of the total modeled footprint for the water-cooled chiller to 96.6\% for the air-cooled chiller. The selection of cooling configuration increased the indirect consumption, shifting the water footprint from the facility site (DCF System) to the generator sites (Energy System).

\begin{table*}[t]
\caption{Annual electricity and water consumption across the modeled cooling and siting cases.}
\label{tab:Ann_E_and_W}

\centering
\footnotesize
\setlength{\tabcolsep}{3pt}
\renewcommand{\arraystretch}{1.15}

\begin{tabular}{p{2.6cm} p{4.1cm} c c c c c}
\toprule

Scenario &
Cooling configuration &
\shortstack{Facility electricity\\(TWh)} &
\shortstack{Direct water\\(million m$^3$)} &
\shortstack{Indirect water\\(million m$^3$)} &
\shortstack{Total water\\(million m$^3$)} &
\shortstack{Indirect share\\(\%)} \\

\midrule

1: Fixed 200 MW IT load&
Water-cooled chiller &
3.08 &
6.45 &
4.48 &
10.93 &
41.0 \\

1: Fixed 200 MW IT load&
Hybrid airside economizer / water-cooled chiller &
2.76 &
3.20 &
3.98 &
7.19 &
55.4 \\

1: Fixed 200 MW IT load&
Air-cooled chiller &
4.06 &
0.21 &
5.91 &
6.12 &
96.6 \\

2: Base locations (concentrated), 4,625 MW build out&
IT liquid cooling: waterside economizer / water-cooled chiller &
36.22 &
70.71 &
53.41 &
124.11 &
43.0 \\

2: Alternative locations, 4,625 MW build out&
IT liquid cooling: waterside economizer / water-cooled chiller &
36.22 &
70.71 &
53.42 &
124.13 &
43.0 \\

\bottomrule

\end{tabular}
\end{table*}

The hybrid configuration further illustrates this tradeoff, shifting between both water-cooled (chiller) and air-cooled (economizer) modes. Its annual direct water consumption (3.20 million m$^3$) fell between the two extremes of the water-cooled and air-cooled configurations, but produced the lowest facility electricity demand (2.76 TWh) and indirect water consumption (3.98 million m$^3$) of all three configurations. Its total water footprint of 7.19 million m$^3$ remained higher than the air-cooled configuration, but was 34\% lower than the water-cooled configuration. The hybrid configuration captured the direct water savings from using the air-cooled mode, without fully absorbing the electricity and indirect water penalties. 

This efficiency gain comes with more complex and climate-dependent operations. Table~\ref{tab:hybrid_modes} shows that the hybrid configuration operated in economizer mode for only 25\% of the year, and the chiller exceeded 80\% of days in the months of January and August through December. The months with the largest economizer share were May and June at 68\% and 53\%, respectively. Figure~\ref{fig:cooling_monthly_water} shows the monthly direct and indirect water consumption for each cooling configuration in Scenario 1. The water--energy tradeoff is clearly shown for the hybrid configuration: compare the water consumption breakdown for May and August. May has among the lowest direct water consumption, given its high economizer use, but higher indirect water consumption. August has high direct water consumption, attributed to higher use of the chiller mode. The operating mode did not result in the expected seasonal pattern of winter-economizer / summer-chiller operations: this is because both temperature and relative humidity conditions had to be satisfied to permit economizer operation. The data center located in Southeast Michigan falls under the IECC 5A climate zone (Cool-Humid), where higher relative humidity restricts how often free-cooling conditions are met, despite lower temperatures.

\begin{table}[h]
\caption{
Monthly operating-mode selection for the hybrid airside economizer /
water-cooled chiller configuration.
}
\label{tab:hybrid_modes}

\centering
\small
\begin{tabular}{lrr}
\toprule
Month & Economizer, Air-Cooled (\%)& Chiller, Water-cooled (\%)\\
\midrule
January   & 10 & 90 \\
February  & 25 & 75 \\
March     & 19 & 81 \\
April     & 33 & 67 \\
May       & 68 & 32 \\
June      & 53 & 47 \\
July      & 32 & 68 \\
August    & 16 & 84 \\
September & 10 & 90 \\
October   & 13 & 87 \\
November  & 17 & 83 \\
December  & 3  & 97 \\
\midrule
Annual    & 25 & 75 \\
\bottomrule
\end{tabular}
\end{table}

\subsection{Average Grid Water Intensities Cannot Capture the True Indirect Water Impacts}\label{sec:result:average_grid_WCI}
\begin{figure}[h]
  \centering
  \includegraphics[width=\linewidth]{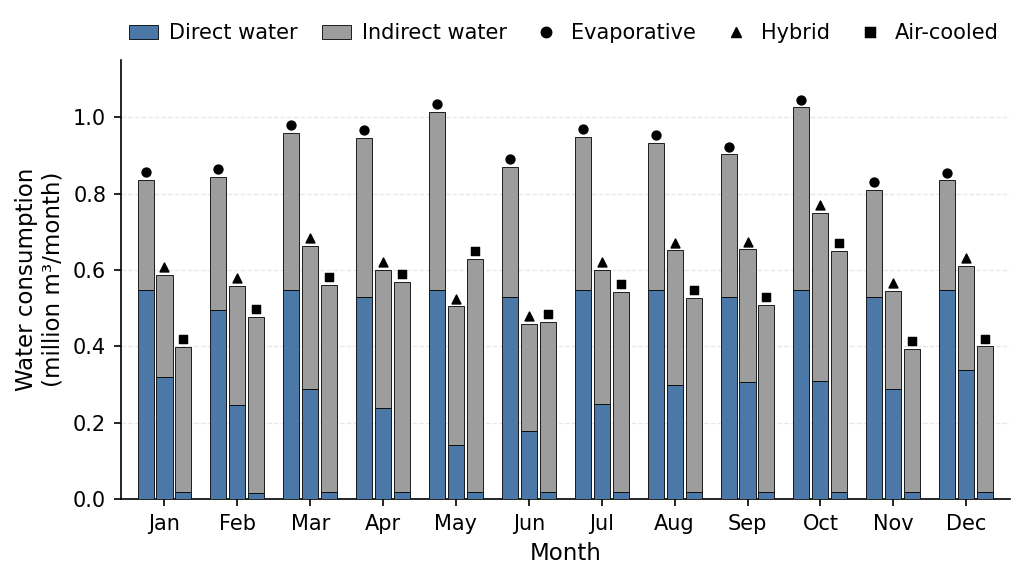}
  \caption{Monthly direct and indirect water consumption across cooling configurations}
  \Description{
  Monthly stacked-bar plots comparing direct and indirect water
  consumption across cooling configurations for the fixed 200 MW
  IT-load analysis and the proposed 4,625 MW facility build out.
  }
  \label{fig:cooling_monthly_water}
\end{figure}

\begin{figure}[h]
  \centering
  \includegraphics[width=\linewidth]{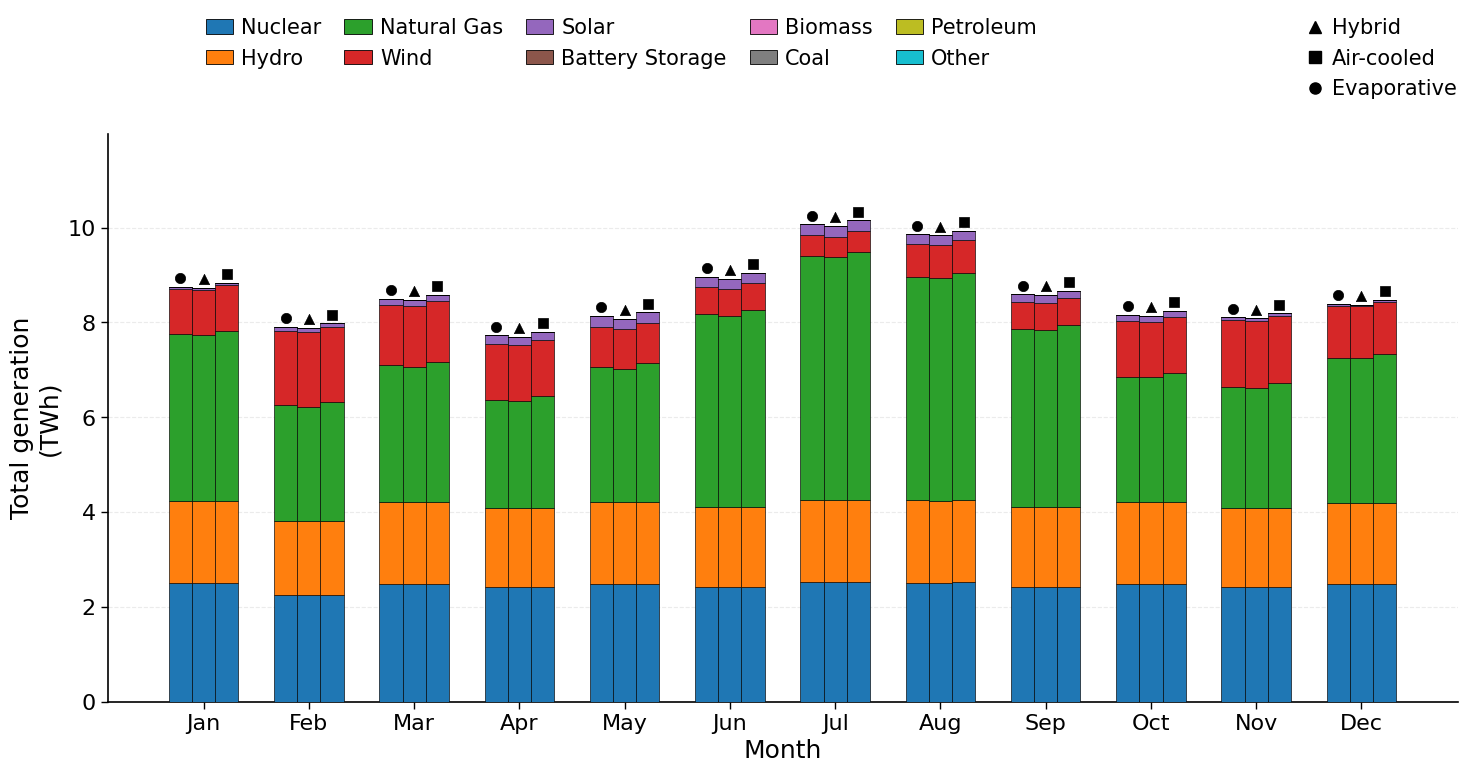}
  \caption{Monthly total generation mix across the three cooling configurations in Scenario 1}
  \Description{
  Monthly total electricity generation by generation technology across the hybrid, air-cooled, and evaporative cooling configurations.
  }
  \label{fig:total_gen_mix}
\end{figure}

Figure~\ref{fig:cooling_monthly_water} shows the monthly direct and indirect water consumption for each cooling configuration in Scenario 1. For the water-cooled (evaporative) and air-cooled configurations, the facility's IT load is fixed throughout the year, resulting in constant monthly direct water consumption (via the constant WUE, where minor variations month-to-month are attributed to the length of the month) and constant monthly power consumption (via the constant PUE). The indirect water consumption, however, varies substantially throughout the year. For example, the indirect water use for the water-cooled configuration is substantially higher in July than in June. This monthly variation does not come from variable power demand from the facility, but is driven by variation in (i) generation availability and (ii) baseline electricity demand. Figure~\ref{fig:total_gen_mix} plots the generation resources dispatched to serve total electric load on the system (baseline + facility load). Nuclear and hydro resources provide baseload throughout the year, operating near full capacity. Wind and solar resources are dispatched when available (intermittent zero-marginal cost generators), and natural gas provides load-following capabilities.

\subsection{Data Center Impacts Extend Beyond the Facility Site} \label{sec:result:spatial_impacts}
The demonstrated water--energy tradeoff has spatial implications: while direct water use impacts subbasins at the facility site, the indirect water use is determined by the generators responding to the added load. In this way, cooling configuration selection also changes the spatial impacts of data center development. 

\begin{figure}[h]
  \centering
  \includegraphics[width=\linewidth]
  {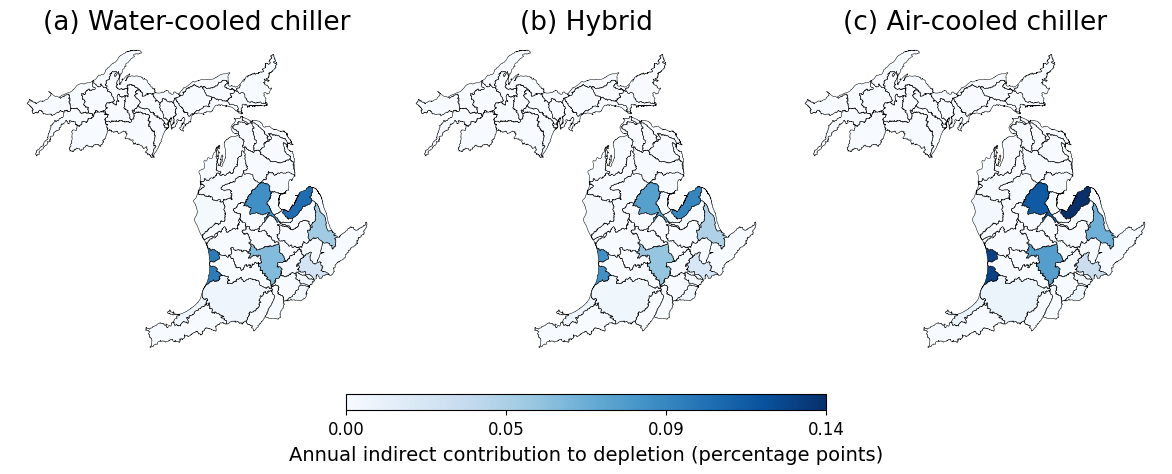}
  \caption{
  Spatial distribution of indirect water impacts across cooling configurations
  }
  \Description{
  HUC-8 maps showing the annual indirect contribution to depletion
  for the water-cooled chiller, hybrid airside economizer /
  water-cooled chiller, and air-cooled chiller configurations.
  All panels use the same color scale.
  }
  \label{fig:cooling_indirect_spatial}
\end{figure}

In Scenario 1, the location of the facility remains the same, so changes in direct water consumption impact a single subbasin. However, when reducing direct water consumption (e.g., switching to air-cooled systems), the corresponding increase in electricity demand increases the magnitude of indirect water consumption and depletion in other subbasins. Figure~\ref{fig:cooling_indirect_spatial} compares the annual indirect depletion impacts across the three cooling configurations. The spatial distribution of indirect depletion remained largely consistent across the cooling configurations. This is likely because of the highly aggregated grid model (only 4 buses and 4 transmission lines) used to represent the Michigan grid. The low spatial resolution of the grid infrastructure, despite highly granular data on the generators and their water intensities, limits the ability to model variable patterns in grid congestion and the resulting generator dispatch which would drive spatial variation. However, the magnitude of impact within those subbasins varied substantially, with the air-cooled configuration producing the largest indirect depletion contributions and the hybrid configuration producing the smallest.

\begin{figure}[h]
  \centering
  \includegraphics[width=\linewidth]{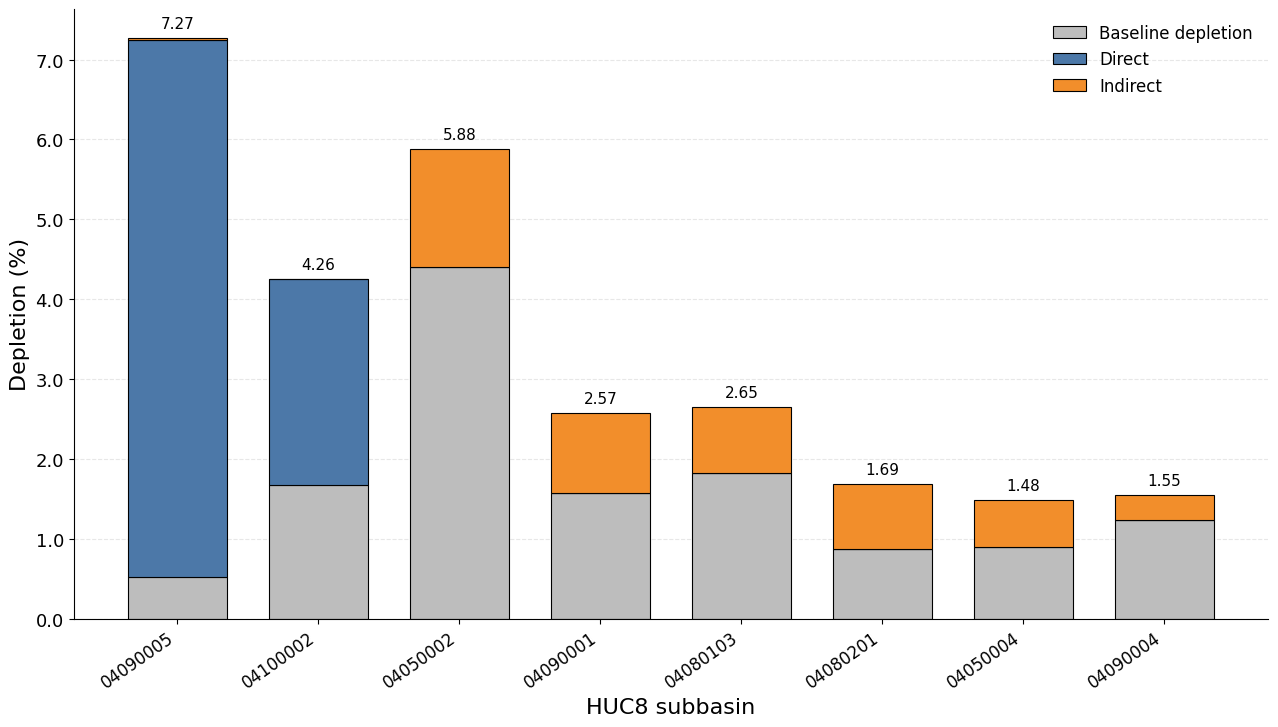}
  \caption{Annual depletion components for the eight HUC-8 subbasins with the largest depletion increases under the proposed data center siting scenario}
  \Description{
  Annual depletion for the eight HUC-8 subbasins with the largest modeled depletion increases under the proposed data center siting scenario.
  }
  \label{fig:huc_annual_depletion}
\end{figure}

\begin{figure}[h]
  \centering
  \includegraphics[width=\linewidth]{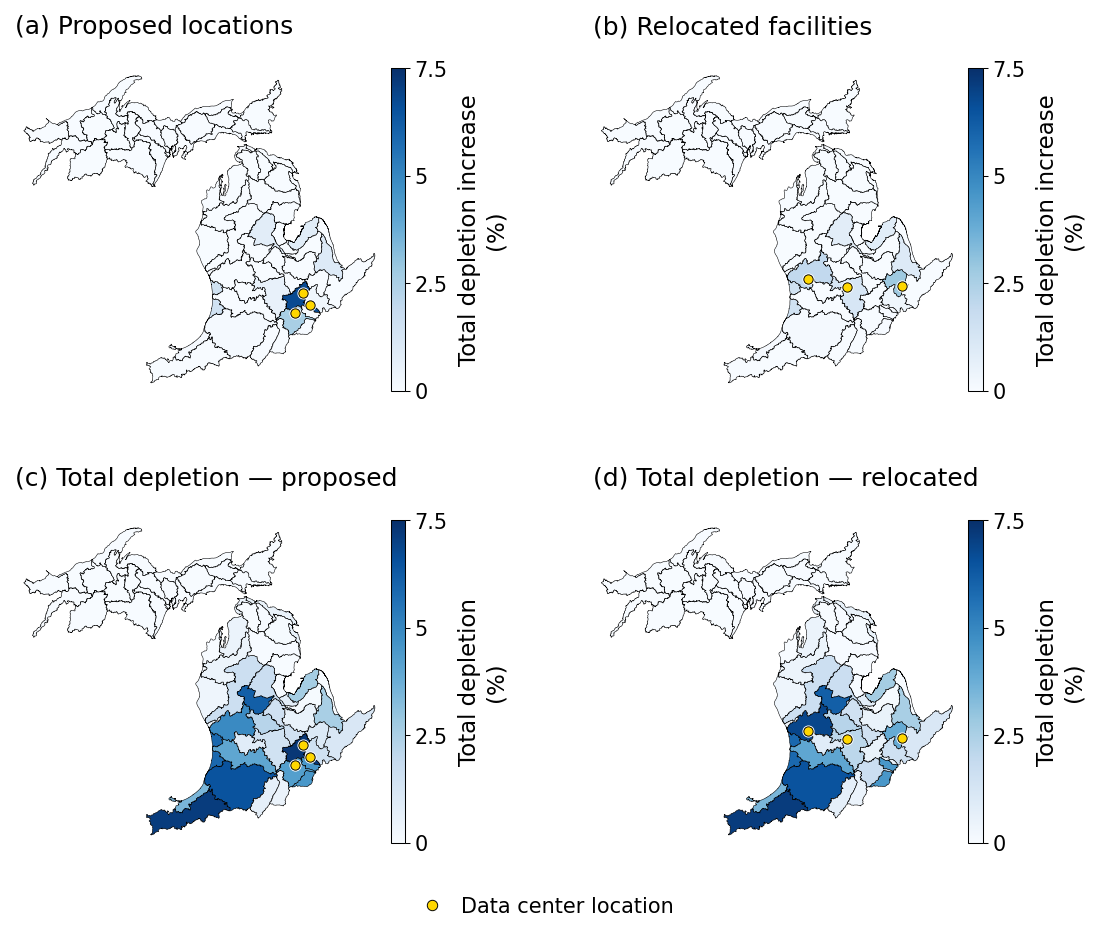}
  \caption{Scenario 2: Spatial distribution of increase in water depletion (top: a and b), and total depletion (bottom: c and d), for the base locations (left) and alternate locations (right). Facility locations are indicated by yellow dots.}
  \Description{
  HUC-8 maps showing the modeled percent change in water depletion
  associated with direct facility water consumption and indirect
  generator water consumption. Yellow points identify proposed
  data center locations.
  }
  \label{fig:spatial_water_depletion}
\end{figure}

In Scenario 2, the water impact of the proposed data center build out is distributed spatially throughout the state. Figure~\ref{fig:huc_annual_depletion} shows the annual depletion at the subbasins with the largest absolute increase in depletion, denoted by their HUC-8 IDs. The baseline (grey), direct (blue), and indirect (yellow) depletion are tracked, showing the magnitude of impact from the data center. Under the concentrated siting case (base case), the three facilities were assigned to two subbasins in Southeast Michigan, producing subbasin depletion increases of 6.74\% and 2.58\%, respectively. For these two subbasins, indirect consumption contributed less than 0.03\% to the total annual increase. The remaining six subbasins experience increased depletion from indirect water consumption. Of these, the top four subbasins experience increases in annual depletion ranging from 0.81\% to 1.5\%, and are located throughout the state, in Southwest, Southeast, North and Central Michigan. While the largest impacts remain concentrated around the proposed facilities, direct water consumption contributes 57\% of the water footprint (see Table~\ref{tab:Ann_E_and_W}), while the remaining 43\% of water consumption is driven by the Energy System and distributed across generator subbasins elsewhere in the state. Methodologies such as those proposed by \cite{wu25} misattribute this consumption to the facility site. The spatial distribution of total depletion increase is shown in Figure~\ref{fig:spatial_water_depletion}(a) and (b) for the base siting and relocation, respectively. Under the relocated case, direct consumption is distributed among three HUC-8 subbasins in Western, Central, and Southeastern Michigan. Although both cases have the same direct water consumption, the distribution of the depletion across different locations reduced the impact of the data center build out: the largest total depletion change in any subbasin decreased to 2.78\% (from 6.74\%).

\subsection{Annualized Reporting Obscures the Underlying System Conditions} \label{sec:results:annual_reporting}
\begin{figure}[h]
  \centering
  \includegraphics[width=\linewidth]{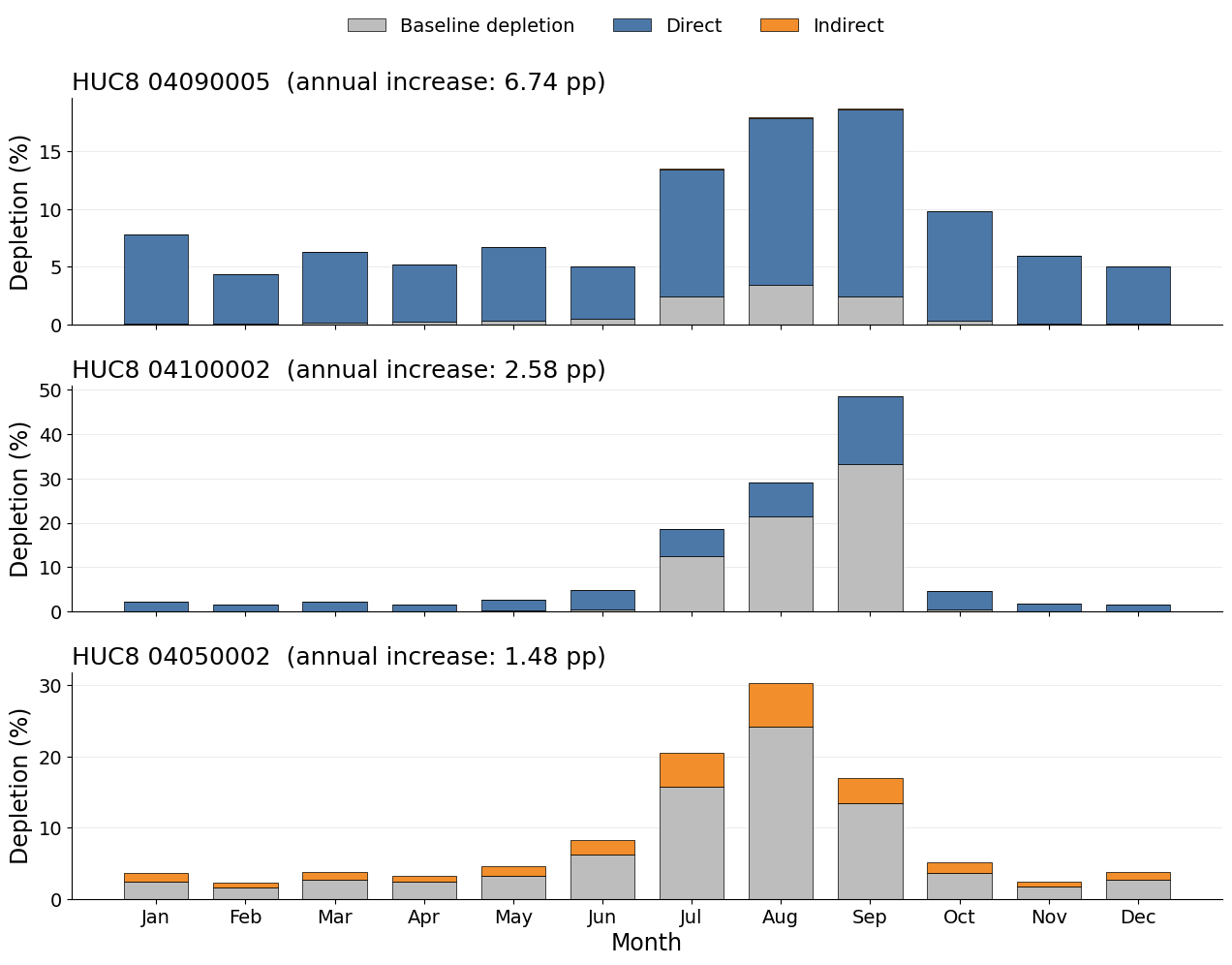}
  \caption{Monthly HUC-8 depletion components for the three most impacted subbasins under the proposed data center siting scenario}
  \Description{
  Monthly depletion for the three HUC-8 subbasins with the largest annual depletion increases under the proposed data center siting scenario.
  }
  \label{fig:huc_breakdown}
\end{figure}

Figure~\ref{fig:huc_breakdown} provides more granular tracking of the water consumption from Scenario 2 (base siting). The plot shows monthly depletion components for the three subbasins with the largest annual depletion increases. Depletion in each of these subbasins varies substantially month to month, driven by water availability and its interactions with the following factors: (i) baseline depletion varies with existing consumption; (ii) direct consumption varies with the load profile; and (iii) indirect consumption varies with generation availability and baseline electricity demand (see Section~\ref{sec:result:average_grid_WCI}). 

Overall, depletion percentage is highest during the summer months due to lower water availability. The two subbasins whose annual increase is dominated by direct consumption (top two plots) show elevated depletion for most of the year, consistent with continuous facility water use. The third subbasin, whose annual increase is entirely from indirect consumption (bottom plot), behaves differently: depletion rises in the late summer and early fall months of July, August, and September, when water availability drops, baseline electric load peaks, and the marginal generators have a higher water cost. Annualized totals for this subbasin understate depletion during these months of high water risk, and overstate depletion in other months.

\subsection{Water Consumption Magnitude Alone Cannot Determine Water Risk} \label{sec:results:water_risk}
Figure~\ref{fig:spatial_water_depletion} compares the increase in depletion from the data center build out (a and b) to the resulting total depletion once that increase is added to existing baseline conditions (c and d). The same information is shown for the eight most impacted subbasins in Figure~\ref{fig:huc_annual_depletion}. The comparison of total increase vs. total depletion shows a mischaracterization of impact when only water consumption magnitude (or depletion increase) is considered. A large increase in depletion may not result in a significant impact (e.g., second-ranked on depletion increase, HUC-8 04100002), while a small increase in depletion may result in a significant impact (e.g., third-ranked on depletion increase, but second-ranked on total depletion, HUC-8 04050002). This latter subbasin -- which experiences a modest 1.48\% increase in depletion -- already has a higher baseline depletion of 4.4\% compared to other subbasins. Further, this increase in depletion occurred entirely from indirect consumption: the facility is not located near this subbasin, and the consumption is entirely from the response of the electricity system to the added load.

\section{Implications for Sustainable Data Center Development}

The results demonstrate that data center water impacts arise from interactions between cooling configuration, electricity demand, grid dispatch, and existing hydrologic conditions. Cooling configuration determines direct consumption and facility electricity demand; grid dispatch determines the magnitude and location of indirect consumption; and existing subbasin conditions determine how that consumption translates to water depletion and risk. An assessment that does not capture all three systems therefore obscures both the magnitude and location of a data center's water footprint. 

\subsection{Recommendations for Industry and Policy}
Grounded in engineering modeling and real-world data, the case study results translate into specific recommendations for industry, grid planning, and reporting practice:

\paragraph{1. Cooling configuration results in a water--energy tradeoff:} Because PUE and WUE assumptions govern the magnitude of water--energy tradeoffs and their seasonal shifts, any impact assessment that assumes annualized fixed values may misrepresent a facility's water and energy demands, and its seasonal variation. In Scenario 1, the air-cooled chiller reduced total water consumption, but its greater electricity demand increased indirect consumption enough to offset part of that saving; the size of the offset depends on the generators supplying the additional electricity, so the same cooling configuration can produce different water outcomes in a different grid region. The range of PUE and WUE values per cooling configuration \cite{shehabi24}, and the potential for climate-dependent operation, further highlights the limitations of generalized facility-level and cooling configuration-type metrics. Without better industry-reported data on specific cooling technologies and their operations, water–-energy assessments will continue to substitute broad performance ranges for real facility behavior. This gap becomes more consequential as hybrid and liquid-cooled systems become more prevalent for AI training infrastructure.

\paragraph{2. Average grid water intensities cannot capture the true indirect water impacts:} Indirect water intensity varies throughout the year, with different generators responding to the added load. A regional or averaged water intensity factor assigns a single water cost to all loads on the system, obscuring the true marginal water cost of adding a unit of load. To accurately capture these impacts, integrated models 
are needed to resolve which generators respond to the added demand, and their corresponding water intensities.

\paragraph{3. Data center impacts extend beyond the facility site:} 

Planning and siting of new developments must consider water conditions at proposed facility sites and in the hydrologic regions where the added electricity generation may increase consumption. The relocation results show that siting can reduce the concentration of direct depletion impacts without reducing total consumption, while the distribution of indirect impacts follows the grid response rather than the facility location. Distributing data center compute is therefore not automatically a better outcome, since the affected subbasins carry different baseline conditions. Evaluating direct and indirect consumption together, alongside those baseline conditions, is critical to sustainable data center development.

\paragraph{4. Annualized reporting obscures the underlying system conditions:} Capturing the temporal variability in system conditions requires data and reporting of both consumption and water availability at sub-annual resolutions. Neither facility-level nor grid-average assessments provide sufficient visibility into the spatio-temporal variations of system conditions to support sustainable data center development. Reporting and impact assessments should require sub-annual, subbasin-level depletion metrics rather than annualized, system-wide totals.

\paragraph{5. Water consumption magnitude alone cannot determine water risk:} Consumption magnitude and the increase in depletion it produces cannot substitute for water risk. Risk depends on how added consumption interacts with existing water uses and availability in a given subbasin -- a function of existing infrastructure, competing consumption, and hydrologic conditions. Reporting frameworks for data center water impact should track total depletion, not just consumption volume or its incremental change. \newline 

\textbf{In summary, facility-level analysis and consumption volumes cannot accurately capture data center impacts. Impact assessments and tracking must move towards integrated water–-energy models that can quantify impacts at a high spatial and temporal granularity, across interconnected infrastructure and natural systems.}

\subsection{Model Limitations and Future Work}
The integrated water--energy model presented is the first to connect facility cooling configuration, grid dispatch, and subbasin-level water depletion for assessing data center sustainability. The model is intentionally modular, and each of its three systems can be extended to capture additional details as better data, higher-resolution models, and industry practice become available.

DCF System: Facility assumptions introduce additional simplification. The DOE PUE and WUE values used represent estimated ranges rather than the true values of specific facilities. Further, the operational model was limited in defining operations-varying PUE and WUE values. For example, the hybrid configuration in Scenario 1 used fixed outdoor temperature and humidity thresholds. This significantly simplifies controls that can be found in systems, including the ability to mix outdoor cold, high humidity air with return-air to enable higher airside economizer utilization in the winter. Second, while the facility load profiles used in Scenario 2 are meant to be representative of an AI hyperscale data center, the load pattern does not reflect different data center uses: AI training, inference, or general data I/O, can have substantially different load shapes and utilization patterns. Incorporating representative, workload-specific load profiles is a topic of future work.

Energy System: The current grid representation aggregates Michigan into four regions, which limits the spatial resolution at which indirect water consumption can be assigned to individual generators and limits the ability to evaluate the impact of transmission congestion. Higher-resolution grid modeling, capable of resolving congestion at finer spatial granularity and generator-level dispatch at finer temporal granularity (e.g., sub-hourly dispatch), is the subject of ongoing work. 

Water System: The hydrological regions are modeled at the subbasin (HUC-8) level. Aggregating the consumption data from HUC-12 to larger HUC-8 regions may average out local differences in water availability and consumption. The resulting depletion estimates indicate regional patterns; more detailed accounting of local water supply and use is needed to assess the risks to specific water regions and sources. Second, assigning a facility or generator to a subbasin using location and distance may misallocate consumption if the water is actually sourced from a different subbasin. Additional data mapping water sources to commercial and municipal users is necessary to accurately capture these local impacts.

General data limitations: Outdated reporting required the analysis to combine electricity, weather, generator, and hydrologic datasets from different years, and gaps in reported generator water use required substitutions that may not reflect individual generator performance. More complete and consistent data collection across these sources would directly improve the model's estimates and further strengthen their applicability to real-world conditions.

\section{Conclusion}

Assessing the water--energy impacts of expanding data center infrastructure requires accounting for both direct water consumption at the facility and indirect water consumption from electricity generation. This study presents an integrated data center--energy--water model that links facility cooling, grid dispatch, and subbasin depletion. The results show that cooling configuration shifts water consumption from the facility to electricity generation, that indirect water intensity varies monthly as the responding generators change, and that facility siting redistributes impacts across the region. Together, these findings demonstrate that energy and water impacts cannot be assessed in isolation, and that sustainable data center development requires integrated water--energy models to assess impacts and support decision making in siting, design, and reporting practices.

\begin{acks}
This work is supported by the University of Michigan's Graham Sustainability Institute Catalyst Grant.
\end{acks}

\bibliographystyle{ACM-Reference-Format}
\bibliography{dews-bibtex}

\newpage 
\appendix

\setcounter{table}{0}
\renewcommand{\thetable}{\Alph{section}.\arabic{table}} 

\section{Generator and Water Consumption Intensity Data} \label{app:mi_gen_data}

Table~\ref{tab:michigan_capacity_mix} reports the modeled installed generating capacity derived from the 2024 EIA-860 plant and generator datasets \cite{eia24a}.

\begin{table}[H]
  \caption{Modeled Michigan installed generating capacity.}
  \label{tab:michigan_capacity_mix}
  \centering
  \small
  \begin{tabular}{lrrr}
    \toprule
    Carrier & Capacity (MW) & Plants & Share (\%) \\
    \midrule
    Natural gas     & 14,445.7 & 58 & 44.6 \\
    Coal            &  6,306.9 &  5 & 19.5 \\
    Wind            &  3,777.3 & 34 & 11.7 \\
    Nuclear         &  3,502.3 &  2 & 10.8 \\
    Hydroelectric   &  2,327.1 & 52 &  7.2 \\
    Solar           &  1,159.1 & 61 &  3.6 \\
    Biomass         &    479.8 & 31 &  1.5 \\
    Petroleum       &    410.2 & 21 &  1.3 \\
    Battery storage &      1.4 &  2 &  0.0 \\
    \midrule
    Total           & 32,409.8 & 266 & 100.0 \\
    \bottomrule
  \end{tabular}
\end{table}

Table~\ref{tab:michigan_marginal_costs} summarizes the marginal cost assumptions used in the economic dispatch model for each grouped generating technology.

\begin{table}[H]
  \caption{Carrier-level marginal-cost assumptions used for dispatch.}
  \label{tab:michigan_marginal_costs}
  \centering
  \small
  \begin{tabular}{lr}
    \toprule
    Carrier & Cost (\$/MWh) \\
    \midrule
    Wind and solar     & 0  \\
    Hydroelectric      & 15 \\
    Nuclear            & 23 \\
    Natural gas        & 23 \\
    Coal               & 41 \\
    Biomass            & 45 \\
    Other generation   & 60 \\
    Petroleum          & 90 \\
    \bottomrule
  \end{tabular}
\end{table}

For reference, Table~\ref{tab:grouped_annual_wci} summarizes the annual water consumption intensity values across the grouped generating technologies represented in the model. Each value is the arithmetic mean of the annual generator WCI values within that technology group.

\begin{table}[H]
  \caption{Mean annual WCI by grouped generating technology.}
  \label{tab:grouped_annual_wci}
  \centering
  \small
  \begin{tabular}{lr}
    \toprule
    Grouped generating technology & WCI (gal/MWh) \\
    \midrule
    Petroleum       & 831.65 \\
    Nuclear         & 671.33 \\
    Coal            & 523.95 \\
    Biomass         & 336.00 \\
    Natural gas     & 193.49 \\
    Solar           &  26.00 \\
    Hydroelectric   &   0.00 \\
    Battery storage &   0.00 \\
    Wind            &   0.00 \\
    \bottomrule
  \end{tabular}
\end{table}
\end{document}